\documentclass[
reprint,
 amsmath,
 amssymb,
 aip
 ]{revtex4-2}

\usepackage{graphicx}
\usepackage{mathtools}

\begin{document}

\title{Optimal control for phase locking of synchronized oscillator populations via dynamical reduction techniques}

\author{Narumi Fujii}
\affiliation{Department of Systems and Control Engineering, Institute of Science Tokyo, Tokyo 152-8552, Japan}
\email{fujii.n.801e@m.isct.ac.jp}
\author{Hiroya Nakao}
\affiliation{Department of Systems and Control Engineering, Institute of Science Tokyo, Tokyo 152-8552, Japan}
\affiliation{Research Center for Autonomous Systems Materialogy, Institute of Science Tokyo, Kanagawa 226-8501, Japan}

\begin{abstract}
We present a framework for controlling the collective phase of a system of coupled oscillators described by the Kuramoto model under the influence of a periodic external input by combining the methods of dynamical reduction and optimal control.
We employ the Ott-Antonsen ansatz and phase-amplitude reduction theory to derive a pair of one-dimensional equations for the collective phase and amplitude of mutually synchronized oscillators.
We then use optimal control theory to derive the optimal input for controlling the collective phase based on the phase equation
and evaluate the effect of the control input on the degree of mutual synchrony using the amplitude equation.
We set up an optimal control problem for the system to quickly resynchronize with the periodic input after a sudden phase shift 
in the periodic input, a situation similar to jet lag, and demonstrate the validity of the framework through numerical simulations.
\end{abstract}

\maketitle

{\bf 
Synchronization of collective oscillations is crucial in biological systems, such as circadian rhythms.
As the loss of synchronization due to disturbances may impair their functionality, e.g., discomfort from jet lag,
a quick recovery to the appropriate synchronized state is desirable.
Like circadian rhythms generated by many neurons, collective oscillations in real-world systems often involve a large number of elements and are difficult to control due to their high dimensionality.
In this study, we present a framework for controlling the collective phase of synchronized oscillator populations using dynamical reduction techniques and optimal control, which enables control of the oscillator population through low-dimensional reduced dynamics.
}

\section{Introduction}

Synchronization of coupled oscillators often serves important functions in biological or artificial systems~\cite{Winfree_1980, glass1988clocks,  Strogatz_1993, stefanovska1999physics, Pikovsky_2001, Kuramoto_2003, ermentrout2010mathematical, Yamaguchi_2003, Golombek_2010, Karma_2013,  Dirk_2012, Marco_2009}, such as firefly bioluminescence~\cite{Strogatz_1993}, circadian rhythms~\cite{Yamaguchi_2003, Golombek_2010}, heartbeat~\cite{Karma_2013},
synchronous power grids~\cite{Dirk_2012}, and swarm robots~\cite{Marco_2009}.
As the loss of synchronization due to disturbances may affect their functionality, 
a quick recovery to the appropriate synchronized state is desirable.

A typical example of such a situation is jet lag.
The circadian rhythm of mammals is generated through collective synchronization of a large number of mutually interacting neurons in the suprachiasmatic nucleus of the brain~\cite{Yamaguchi_2003, Golombek_2010}.
This collective rhythm is further synchronized with (phase-locked to) the periodic light-dark input of the diurnal cycle.
When the synchronization of the circadian rhythm with the diurnal cycle is disturbed by traveling across time zones, it leads to 
jet lag, i.e., sleep disorder causing  drowsiness, headaches, fatigue, etc.
Therefore, finding the control input that facilitates quick recovery of appropriate synchronized states is an important issue.
Motivated by biological or engineering problems, synchronization control has been actively studied, 
with many of the studies focusing on the phase oscillators as the control target
~\cite{Moehlis_2006a, Moehlis_2006b, Zlotnik_2013, Jr-Shin_2016, Tanaka_2008, Kawamura_2008, Monga_2019, Takata_2021, Kato_2021, Ozawa_2021, Berman_2022, Petar_2023, Yawata_2024, Wilson_2024, Namura_2024a, Namura_2024b}.

The Kuramoto model is a paradigmatic mathematical model that describes the phase dynamics of a population of globally coupled self-sustained oscillators exhibiting 
collective synchronization~\cite{Kuramoto_2003, Strogatz_2000}.
However, since synchronization control of the Kuramoto model is a nonlinear control problem, 
it is difficult to handle realistic high-dimensional situations with many oscillators. 
Also, it is generally impossible to control the individual oscillators in the population, and only a global control signal common to all the oscillators can be applied~\cite{Jr-Shin_2016, Kawamura_2008}.
Therefore, dynamical reduction techniques to derive low-dimensional description of the high-dimensional coupled oscillators have been developed~\cite{Kuramoto_2003,Kawamura_2008,Ott_Antonsen_2008,Kawamura_2010a,Kawamura_2010b}.

In the theory of coupled oscillators, the continuum limit, where the number of oscillators tends to infinity, 
is often analyzed~\cite{Kuramoto_2003}. 
In Ref.~\onlinecite{Kawamura_2008}, phase reduction of the nonlinear Fokker-Planck equation describing a population of noisy Kuramoto oscillators in the continuum limit was performed. 
Mutual synchronization between populations of Sakaguchi-Kuramoto-type oscillators was analyzed by using phase reduction of the nonlinear Fokker-Planck equation in Ref.~\onlinecite{Kawamura_2010a},  
and by using the equation for the complex order parameter obtained by the Ott-Antonsen(OA) ansatz in Ref.~\onlinecite{Kawamura_2010b}, respectively.
Various models, including periodically driven systems,
have also been analyzed using the OA ansatz in Refs.~\onlinecite{Wolfrum_2013, Strogatz_2008, Ernest_2014, Pikovsky_2019}. 
Regarding control of oscillator populations,
the nonlinear Fokker-Planck equation was used to design the control input in Ref.~\onlinecite{Berman_2022}, 
and a comprehensive phase diagram was obtained for the Sakaguchi-Kuramoto model with feedback control terms by using the OA ansatz
in Ref.~\onlinecite{Ozawa_2021}.

In this paper, we develop a framework for controlling the synchronization of high-dimensional Kuramoto-type coupled oscillators with a periodic input by combining the methods of dynamical reduction and nonlinear optimal control~\cite{Lewis_2012, Kirk_2004}.
We employ the OA ansatz~\cite{Ott_Antonsen_2008} and phase-amplitude reduction~\cite{Wilson_2016,Shirasaka_2017} to derive a pair of phase and amplitude equations describing the synchronized collective oscillations of the oscillators in the continuum limit.
The derived phase equation is then used for designing the optimal control input, and the amplitude equation is used for evaluating the effect of the control input on the degree of mutual synchrony of the oscillators.
Considering a situation like jet lag, we analyze a problem of controlling the collective phase of the oscillators to quickly resynchronize with the periodic external input, after a sudden phase shift in the periodic input caused, e.g., by traveling across timezones.

This paper is organized as follows.
First, we introduce the Kuramoto model under a periodic input and numerically simulate the loss of synchronization with the periodic input due to a sudden shift in the phase of the periodic input in Sec.~II.
Next, we reduce the dynamics of the Kuramoto model to the dynamics of a complex order parameter using the OA ansatz, and further reduce it to the phase and amplitude equations in Sec.~III.
Then, we formulate the optimal control problem using the reduced phase equation in Sec.IV, present the results of numerical simulations, and evaluate the effect of control input on the system in Sec.~V.
Finally, Sec.~VI. gives the conclusion.

\section{Model}

\subsection{Kuramoto model with a periodic input}

The Kuramoto model describes a system of phase oscillators with all-to-all coupling~\cite{Kuramoto_2003}. We consider a Kuramoto model subjected to a sinusoidal periodic input, described by
\begin{align}
        \dot{\phi}_j = \omega_j + \frac{K}{N} \sum_{k=1}^{N} \sin(\phi_k - \phi_j)+ z(\phi_j)p_0\sin\Theta(t),~~\notag\\(j=1,\cdots,N),
        \label{eq:Kuramoto_model_external}
\end{align}
where $\phi_j$ and $\omega_j$ are the phase and the natural frequency of the $j$-th oscillator, respectively, $K$ is the coupling strength, $N$ is the number of oscillators,
$z(\phi_j)$ is 
the individual phase sensitivity function 
(a.k.a. infinitesimal phase response function / phase resetting curve) 
characterizing the linear response of each oscillator to applied perturbations~\cite{Kuramoto_2003,Nakao_2016}, $p_0$ is the amplitude of the periodic input, and $\Theta(t)$ is the phase of the periodic input that increases with a constant frequency $\omega_p$.
We assume that the system is high-dimensional $(N \gg 1)$, the amplitude $p_0$ of the input is sufficiently small, 
and $\omega_j$ follows a Lorentz distribution,
\begin{align}
        g(\omega) = \frac{\gamma}{\pi} \frac{1}{\left(\omega - \omega_0\right)^2 + \gamma^2}, 
        \label{eq:Lorentz_distribution} 
\end{align}
where $\omega_0$ is the central frequency and $\gamma$ is the width parameter.
We assume that the input frequency $\omega_p$ is close to $\omega_0$,
and that the individual phase sensitivity function is given by a simple sinusoidal form, $z(\phi_j)=\sin\phi_j$.
The above Kuramoto model with the lowest-harmonic sinusoidal individual phase sensitivity function can be obtained by phase reduction of a system of globally coupled Stuart-Landau oscillators driven by a periodic input~\cite{Kuramoto_2003,Nakao_2016}.

The complex order parameter characterizing the degree of mutual synchrony of the system, which we also call the complex mean field, is defined as~\cite{Kuramoto_2003}
\begin{align}
        A(t) = R(t)e^{i\Phi(t)} = 
        \frac{1}{N} \sum_{j=1}^{N} 
        e^{i \phi_j(t)}.
        \label{eq:Order_parameter}
\end{align}
Here, $i=\sqrt{-1}$ is the imaginary unit, the modulus $R$ of $A$ 
gives the amplitude of the complex mean field characterizing the degree of mutual synchrony, and the argument $\Phi$ of $A$ represents the collective phase of the complex mean field, i.e., the collective phase of the system;
$R$ takes a value in $[0, 1]$, where $R=0$ when the oscillators are incoherent and $R=1$ when the oscillators are completely synchronized.

By using $R(t)$ and $\Phi(t)$, we can rewrite the interaction term in Eq.~(\ref{eq:Kuramoto_model_external}) as
\begin{align}
        \dot{\phi}_j = \omega_j + K R(t) \sin(\Phi(t) - \phi_j)+ z(\phi_j)p_0\sin\Theta(t).
\end{align}
Thus, we can represent Eq.~(\ref{eq:Kuramoto_model_external}) formally as a one-body problem driven by the complex mean field and the periodic input.

\subsection{Loss of synchronization with the periodic input}

In this study, we assume that the system is exhibiting synchronized collective oscillations under the periodic input,
and focus on the recovery of the system from a loss of synchronization with the periodic input.
When the system is synchronized with (phase-locked to) the periodic input, the difference $\Psi(t) = \Phi(t) - \Theta(t)$ between the corrective phase $\Phi(t)$ of the system and the phase $\Theta(t)$ of the periodic input is kept small. 
This occurs when the frequency $\omega_p$ of the periodic input is close to $\omega_0$.

We consider a jet lag like situation and introduce a sudden phase shift in the periodic input 
in the steady synchronized state of the system with the periodic input, given by
\begin{align}
        \Theta(t)=\left\{\begin{array}{l}
        \omega_p t \quad (0 \leq t<t_s), \\
        \omega_p t+\Delta \Theta \quad (t \geq t_s),
        \end{array}\right.
        \label{eq:Jet_lag}
\end{align}
where $\Delta \Theta$ is the phase shift of the periodic input and $t_s$ is the time at which the phase shift is applied.

\begin{figure}[tb] 
    \centering
    \includegraphics[width=1.0\linewidth]{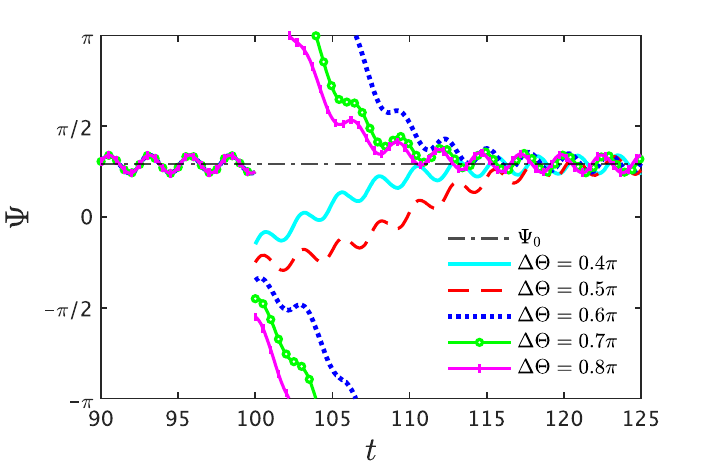}
    \caption{
    Time series of the phase difference $\Psi$. The phase shift is introduced to the periodic input at $t = t_s = 100$.
    The parameters of the Kuramoto model are $K=5.0$, $N=10^4$, $\omega_0=1.0$, and $\gamma=0.5$. 
    The amplitude and frequency of the periodic input are $p_0=0.75$ and  $\omega_p=1.2$, respectively.
    }
    \label{fig:Phase_Different_nocont}
\end{figure}

Figure~\ref{fig:Phase_Different_nocont} shows the time series of the phase difference $\Psi(t)$ for several values of the given phase shift, $\Delta \Theta=0.4\pi,~0.5\pi,~0.6\pi,~0.7\pi,$ and $0.8\pi$, where the phase shift is introduced at $t_s=100$.
Before the phase shift, the phase difference $\Psi(t)$ exhibits small-amplitude oscillations around 
a certain value $\Psi_0$
and does not grow, i.e.,  the collective phase $\Phi(t)$ of the system is locked to the phase $\Theta(t)$ of the external input. 
Note that we observe small-amplitude oscillations of $\Psi(t)$ in the phase-locked state because the effect of the periodic input is not averaged over time in the present model; the phase difference converges to a constant only when further averaging is performed on Eq.~\eqref{eq:Kuramoto_model_external}~\cite{Nakao_2016}.
By the introduction of the phase shift of the periodic input at $t_s$, the phase difference $\Psi(t)$ is suddenly shifted to a different value from $\Psi_0$, and then gradually goes back to $\Psi_0$. This process of resynchronization (recovery of phase locking) with the periodic input takes a certain amount of time.

\begin{figure}[tb] 
    \centering
    \includegraphics[width=1.0\linewidth]{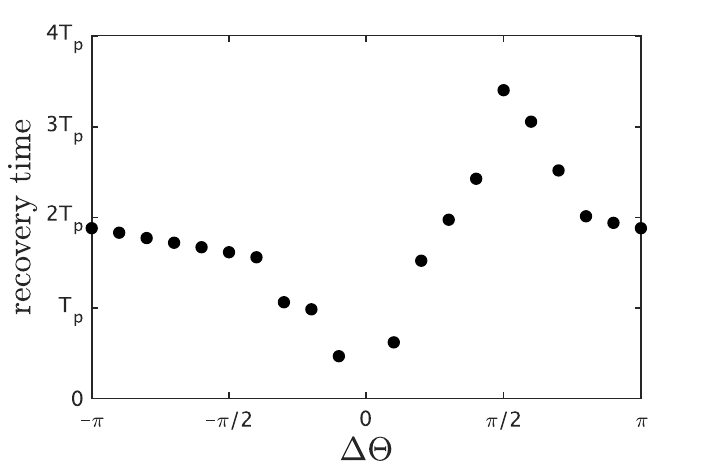}
    \caption{
    Dependence of the recovery time $t_r-t_s$ on the phase shift $\Delta\Theta$. Recovery time is plotted against $\Delta\Theta$ at intervals of $0.1\pi$. $T_p=2\pi/\omega_p$ is the period of the input.}        
    \label{fig:recover_time}
\end{figure}

We define that the time difference $t_r-t_s$ between the time $t_s$ at which the input phase is shifted and the time $t_r$ at which $\Psi(t)$ returns to $\Psi_0$ for the first time after $t_s$ as the recovery time.
The dependence of the recovery time on the phase shift $\Delta\Theta$ is shown in Fig. \ref{fig:recover_time}.
The recovery time takes a peak around $\Delta\Theta=0.5\pi$. In the context of jet lag, this is called the jet lag separatrix, and is considered to be related to the east-west asymmetry in the recovery time from jet lag, as discussed in \cite{Kori_2017,Lu_2016}.

\subsection{Control of resynchronization}

\begin{figure}
    \centering
    \includegraphics[width=1.0\linewidth]{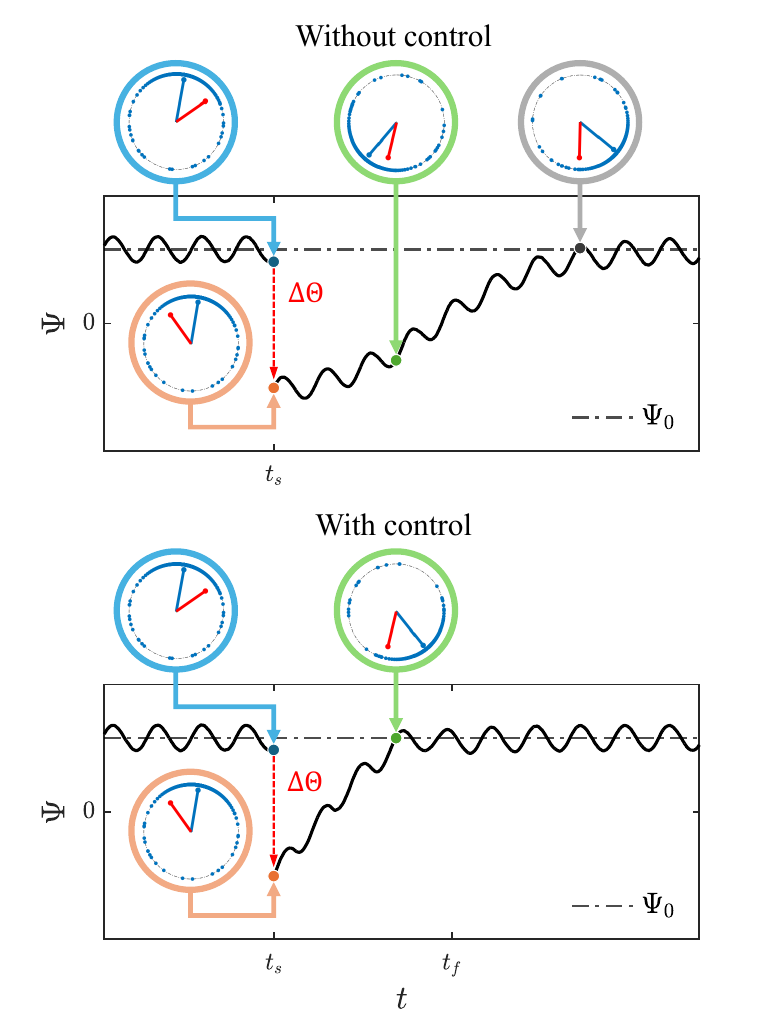}
    \caption{Control of oscillator populations for quick resynchronization with a periodic input.
    It takes a certain amount of time for the collective phase (blue line) to resynchronize with 
    the phase of the periodic input (red line) after a given phase shift. 
    The control input is designed over the interval $[t_s, t_f]$ to achieve quick resynchronization.}
    \label{fig:schematic}
\end{figure}

In this study, we use optimal control theory to derive the control input that realizes quick recovery from the loss of synchrony with the periodic input after the sudden phase shift, as shown schematically in Fig. \ref{fig:schematic}.
Since the system is high-dimensional ($N \gg 1$), control of all oscillator phases 
by applying each individual oscillator a different input $u_j(t)$ is impossible.
Therefore, we assume that all oscillators receive a common control input $u(t)$ in addition to the periodic input $p_0 \sin \Theta(t)$ as
\begin{align}
        \dot{\phi}_j = 
        \omega_j &+ K R(t) \sin(\Phi(t) - \phi_j) 
        +z(\phi_j) p(t),
        \label{eq:Kuramoto_model_external_control}
\end{align}
for all $j$, where
\begin{align}
p(t)=p_0\sin\Theta(t)+u(t).
\end{align}
We also assume that the net input $p(t)$ to the system is weak.
That is, we aim to reestablish phase locking with the periodic input using only a weak control input, so that the control input does not disturb the collective synchronization of the oscillators in the system, which may play important functions in the case of biological systems~\cite{Winfree_1980, glass1988clocks, Strogatz_1993}.

As our interest is on the collective phase of 
mutually synchronized oscillators, in the following sections,
we first consider the continuum limit with $N \to \infty$ and derive a low-dimensional reduced equation describing the system's collective dynamics,
and further reduce it to a pair of phase-amplitude equations. We then design $u(t)$ using the optimal control theory.

\section{Dynamical reduction}

In this section, we briefly review the continuum limit of the Kuramoto model and the OA ansatz, and then perform phase-amplitude reduction of the resulting equation for the complex order parameter.

\subsection{Continuum limit}

We first consider the continuum limit of the Kuramoto model with an external input~\cite{Kawamura_2008}.
We introduce the probability density function $P(\phi, \omega, t)$ of the phase $\phi$ and frequency $\omega$ of the oscillators at time $t$.
In the $N \to \infty$ limit, the order parameter $A(t)$ in Eq.~\eqref{eq:Order_parameter} can be expressed as
\begin{align}
        A(t) &= R(t) e^{i \Phi(t)} \notag\\
        &=\int_{0}^{2\pi}\int_{-\infty}^{\infty}e^{i \phi'}
        P(\phi',~\omega',~t)g(\omega')
        d\omega' d\phi'
        \label{eq:Order_parameter_PDF}
\end{align}
by using $P(\phi, \omega, t)$. 
The time evolution of $P(\phi,\omega,t)$ is given by the continuity equation~\cite{Strogatz_2000, Acebron_2005}, which is explicitly given from Eq.~(\ref{eq:Kuramoto_model_external_control}) with the above $R(t)$ and $\Phi(t)$ as
\begin{align}
        \frac{\partial P}{\partial t}&=-\frac{\partial}{\partial \phi}\left(\dot{\phi} P\right)\notag\\
        &=-\frac{\partial}{\partial \phi} \left[ \lbrace
        \omega+ KR(t) \sin\left(\Phi(t)-\phi\right)+p(t)\sin\phi \rbrace P \right].
        \label{eq:Continuity}
\end{align}
Using Eq.~(\ref{eq:Order_parameter_PDF}), Eq.(\ref{eq:Continuity}) is explicitly written as
\begin{align}
    \frac{\partial P}{\partial t} &= -\frac{\partial}{\partial \phi} \Bigg[ \Bigg\{ \omega  + p(t)\sin\phi \notag\\+& K \int_{0}^{2\pi} \int_{-\infty}^{\infty} \sin(\phi' - \phi) P(\phi', \omega', t) g(\omega') d\omega' d\phi' \Bigg\} P \Bigg].
    \label{eq:time_evolution_of_PDF}
\end{align}
This nonlinear continuity equation represents the dynamics of $P(\phi,\omega,t)$ of the system corresponding to Eq.~(\ref{eq:Kuramoto_model_external_control}) in the continuum limit.
Since oscillators with different $\omega$ do not mix with each other, the normalization condition $\int_0^{2\pi} P(\phi, \omega, t) d\phi = 1$ holds independently for each value of $\omega$.

\subsection{Ott-Antonsen ansatz}

Next, we introduce the OA ansatz to reduce Eq.~(\ref{eq:time_evolution_of_PDF}) to the dynamics of the complex order parameter in two dimensions~\cite{Ott_Antonsen_2008}.
We expand $P(\phi,\omega,t)$ in a Fourier series as
\begin{align}
        P(\phi,~\omega,~t)=\frac{1}{2\pi}\sum_{m=-\infty}^{\infty}P_m(\omega,~t)e^{i m\phi}.
        \label{eq:Fourie}
\end{align}
Substituting Eq.~(\ref{eq:Fourie}) into Eq.(\ref{eq:time_evolution_of_PDF}), we obtain 
\begin{align}
        \frac{\partial}{\partial t}P_m& = - i m \omega 
        P_m\notag-\frac{mp(t)}{2}\left(P_{m-1}-P_{m+1}\right)\notag\\
        &+\frac{Km}{2}\Big\{ P_{m-1}\int_{-\infty}^{\infty}P_1(\omega',~t)g(\omega')d\omega'
        \notag\\&-P_{m+1}\int_{-\infty}^{\infty}P_{-1}(\omega',~t)g(\omega')d\omega'\Big\}
        \label{eq:p_m}
\end{align}
for each Fourier coefficient $P_m(\omega, t)$, where $P_m(\omega,~t)=\bar{P}_{-m}(\omega,~t)$ because $P(\phi,~\omega,~t)$ is real (overline represents the complex conjugate).

Ott and Antonsen assumed a special class of $P(\phi, \omega, t)$ characterized by the Fourier coefficients of the form~\cite{Ott_Antonsen_2008}
\begin{align}
        P_m(\omega,~t)=\alpha(\omega,~t)^m,~~(m=1,~2,~\cdots),
        \label{eq:Ott-Antonsen_ansatz}
\end{align}
i.e., those given by powers of a variable $\alpha(\omega, t)$.
Note that $P_0(\omega,~t) = 1$ by the normalization condition for $P(\phi, \omega, t)$.
Plugging this form into Eq.~\eqref{eq:p_m}, we obtain
\begin{align}
        \frac{\partial}{\partial t}\alpha&= - i \omega 
        \alpha-\frac{p(t)}{2}\left(1-\alpha^{2}\right)\notag\\
        &+\frac{K}{2}\Big\{ \int_{-\infty}^{\infty}\alpha(\omega',~t)
        g(\omega')d\omega'\notag\\
        &-\alpha^{2}\int_{-\infty}^{\infty}\alpha(\omega',~t)^{-1}
        g(\omega')d\omega'\Big\}
        \label{eq:a_m1}
\end{align}
for $m=1$.
From Eqs.~(\ref{eq:Order_parameter_PDF}) and (\ref{eq:Ott-Antonsen_ansatz}), the complex order parameter can be expressed as
\begin{align}
        A(t)=\int_{-\infty}^{\infty}\alpha(\omega',~t)^{-1} g(\omega')d\omega',
        \label{eq:Order_parameter_Ott-Antonsen}
\end{align}
and since $P_m=\bar{P}_{-m}$, 
\begin{align}
\alpha(\omega,~t)^m=\overline{\left\{\alpha(\omega,~t)^{-m}\right\}}=
\left\{\overline{\alpha(\omega,~t)}\right\}^{-m}.
\end{align}
Therefore, Eq.~(\ref{eq:a_m1}) is rewritten as
\begin{align}
        \frac{\partial}{\partial t}\alpha = - i \omega 
        \alpha-\frac{p(t)}{2}\left(1-\alpha^{2}\right)
        +\frac{K}{2}\lbrace \overline{A(t)}-\alpha^{2}A(t)\rbrace.
        \label{eq:a}
\end{align}
For the Lorentzian $g(\omega)$ given by Eq.~\eqref{eq:Lorentz_distribution}, we obtain
\begin{align}
        A(t)=\overline{\alpha(\omega_0 - i \gamma,~t)}
\end{align}
by applying the residue theorem in Eq.~\eqref{eq:Order_parameter_Ott-Antonsen}, where
$\alpha(\omega,~t)$ is analytically continued to the complex plane.
Thus, the complex order parameter $A(t)$ is determined by $\alpha(\omega,~t)$ at $\omega=\omega_0 - i \gamma$, and Eq.~\eqref{eq:a} can be expressed by using $A(t)$ as
\begin{align}
        \dot{A}(t)=\left(\frac{K}{2}-\gamma + i \omega_0\right) {A}(t)
        -\frac{K}{2}{A}(t)|A(t)|^2\notag\\-\frac{p(t)}{2}\lbrace1-{A}(t)^{2}\rbrace.
        \label{eq:Dynamics_of_order_parameter}
\end{align}
This equation represents two-dimensional dynamics of the complex order parameter $A(t)$.

When the perturbation $p(t)$ is absent, this model represents the Stuart-Landau oscillator, i.e., the normal form of the supercritical Hopf bifurcation~\cite{Kuramoto_2003,Nakao_2016}, which is analytically solvable.
If $\frac{K}{2}-\gamma>0$, this system with $p(t)=0$ has a stable limit cycle, corresponding to the collective oscillations 
of the system.
In what follows, we assume that the parameters $K$ and $\gamma$ are in the above range and the oscillators are mutually 
synchronous to a certain degree.

\subsection{Phase-amplitude reduction}

Next, we introduce phase-amplitude reduction theory for describing the dynamics of a stable limit cycle under weak perturbation~\cite{Wilson_2016,Shirasaka_2017}. 
Since Eq.~(\ref{eq:Dynamics_of_order_parameter}) is a Stuart-Landau equation with an external input $p(t)$,
we can derive a pair of approximate phase and amplitude equations assuming that $p(t)$ is sufficiently small.
Application of the phase reduction to collective rhythms of globally coupled oscillators via OA ansatz was performed in Ref.~\onlinecite{Kawamura_2010b}, and the following analysis is a generalization to phase-amplitude reduction.

Using the modulus $R(t)$ and argument $\Phi(t)$ of $A(t)$, i.e., $A(t)=R(t)e^{i \Phi(t)}$, 
Eq.~(\ref{eq:Dynamics_of_order_parameter}) without $p(t)$ is rewritten as
\begin{align}
        \dot{R}&=\left(\frac{K}{2}-\gamma\right)R-\frac{K}{2}R^3,
        \label{eq:Dynamics_of_order_parameter_without_input_R}\\
        \dot{\Phi}&=\omega_0.
        \label{eq:Dynamics_of_order_parameter_without_input_Phi}
\end{align}
Under the condition that $\frac{K}{2}-\gamma>0$, 
this system has a stable limit cycle given by
\begin{align}
        R(t) = R_0 := \sqrt{1-\frac{2\gamma}{K}},\quad
        \Phi(t)  = \omega_0 t + const.
        \label{eq:Limit_cycle}
\end{align}
Now let $A(t)=X(t)+iY(t)$ and $\boldsymbol{X}(t)=(X(t),~Y(t))$, where $X$ and $Y$ are the real and imaginary parts of $A$, respectively, and rewrite Eq.~(\ref{eq:Dynamics_of_order_parameter}) as
\begin{align}
        \dot{\boldsymbol{X}}=\boldsymbol{F}(\boldsymbol{X})+\boldsymbol{G}(\boldsymbol{X}) p(t),
        \label{eq:Dynamics_of_order_parameter_with_input_on_X_FG}
\end{align}
where
\begin{align}
        \boldsymbol{F}(\boldsymbol{X})&=\left(\begin{array}{cc}
                \frac{K}{2}-\gamma & -\omega_0 \\
                \omega_0 & \frac{K}{2}-\gamma
                \end{array}\right)\binom{X}{Y}\notag\\
                &-\frac{K}{2}\left(X^2+Y^2\right)\binom{X}{Y},\\
        \boldsymbol{G}(\boldsymbol{X})&=\frac{1}{2}\binom{X^2-Y^2-1}{2 X Y}.    
\end{align}
From Eq.~\eqref{eq:Limit_cycle}, the unperturbed system $\dot{\boldsymbol X} = {\boldsymbol F}({\boldsymbol X})$ has a stable limit-cycle solution ${\boldsymbol X}_0(t) = R_0 ( \cos \omega_0 t,\ \sin \omega_0 t)$, where the initial condition is taken as ${\boldsymbol X}_0(0) = (1, 0)$.
In the basin of this limit cycle, the asymptotic phase function $\Xi(\boldsymbol{X})$ satisfying
\begin{align}
    &\dot{\Xi}(\boldsymbol{X})
    =\text{grad}_{\boldsymbol{X}}\Xi(\boldsymbol{X})\cdot \boldsymbol{F}(\boldsymbol{X})
    =\omega
    \label{eq:phase_function_def}
\end{align}
can be introduced, which increases at a constant frequency 
$\omega$~\cite{Kuramoto_2003,Nakao_2016}.
Similarly, the amplitude function $\Gamma(\boldsymbol{X})$ can also be introduced to satisfy
\begin{align}
    \dot{\Gamma}(\boldsymbol{X})
    &=\text{grad}_{\boldsymbol{X}}\Gamma(\boldsymbol{X})\cdot \boldsymbol{F}(\boldsymbol{X})
    =\mu \Gamma(\boldsymbol{X}),
    \label{eq:amplitude_reduction_def}
\end{align}
where $\mu < 0$ is the Floquet exponent with a negative real part
of the unperturbed system and $\Gamma(\boldsymbol{X}_0(t))=0$ on the limit cycle $\boldsymbol{X}_0(t)$~\cite{Mauroy_2013,Wilson_2016,Shirasaka_2017}. 
The amplitude characterizes the distance from the limit cycle and decreases exponentially.
These functions are essentially the Koopman eigenfunctions~\footnote{The complex exponential $e^{i \Xi({\boldsymbol X)}}$ is a Koopman eigenfunction associated with the eigenvalue $i\omega$. $\Gamma(\boldsymbol X)$ is itself a Koopman eigenfunction with the eigenvalue $\mu$.} of the system associated with the eigenvalues $i \omega$ and $\mu$~\cite{Mauroy_2013,Wilson_2016,Shirasaka_2017}.
As the system is two-dimensional, we have only a single amplitude in addition to the phase.

The phase function and amplitude functions satisfying Eqs.~\eqref{eq:phase_function_def} and \eqref{eq:amplitude_reduction_def} can be explicitly obtained as
\begin{align}
    \Xi(\boldsymbol{X})&= \arctan (Y / X),
    \label{eq:phase_function}\\
    \Gamma(\boldsymbol{X})&=\frac{K}{2}\left(1-\frac{R_0^2}{X^2+Y^2}\right),
    \label{eq:amplitude_function}
\end{align}
where the frequency $\omega$ is equal to $\omega_0$ and the non-zero Floquet exponent is given by $\mu = -KR_0^2$.
Thus, the asymptotic phase $\Xi$ is simply equal to the argument $\Phi$ of the complex order parameter $A$, i.e., the collective phase of the system.
This is because the coefficient of the cubic term in Eq.~\eqref{eq:Dynamics_of_order_parameter} is real.
In what follows, we denote the system state on the limit cycle with phase $\Phi$ as $\boldsymbol{\chi}(\Phi) = \boldsymbol{X}_0(\Phi/\omega)$.

Using Eqs.~\eqref{eq:phase_function} and \eqref{eq:amplitude_function}, we can derive approximate phase and amplitude equations 
from the weakly perturbed dynamics, Eq.~(\ref{eq:Dynamics_of_order_parameter_with_input_on_X_FG})~\cite{Wilson_2016,Shirasaka_2017}.
First, the equation for the collective phase $\varphi = {\Xi}(\boldsymbol{X})$ of the system can be obtained as
\begin{align}
    \dot{\varphi} &=\omega_0+\text{grad}_{\boldsymbol{X}= \boldsymbol{X}(t)}\Xi(\boldsymbol{X})\cdot 
    \boldsymbol{G}(\boldsymbol{X}(t)) p(t)\notag\\
    &\simeq \omega_0+\text{grad}_{\boldsymbol{X}= \boldsymbol{\chi}(\varphi(t))}\Xi(\boldsymbol{X})\cdot 
    \boldsymbol{G}\left(\boldsymbol{\chi}(\varphi(t))\right) p(t)\notag\\
    &=\omega_0+\boldsymbol{Z}(\varphi(t))\cdot\boldsymbol{G}(\varphi(t)) p(t).
    \label{eq:phase_reduction}
\end{align}
Here, the second term is approximately evaluated at $\boldsymbol{X} = \boldsymbol{\chi}(\varphi(t))$ 
on the limit cycle, assuming that the external input is sufficiently small and the system state always stays near the limit cycle,
and the phase sensitivity function $\boldsymbol{Z}(\varphi)=\text{grad}_{\boldsymbol{X}= \boldsymbol{\chi}(\varphi)}\Xi(\boldsymbol{X})$,
the gradient of the asymptotic phase function on the limit cycle at phase $\varphi$,
is introduced. 
Next, the equation for the amplitude $r = \Gamma({\boldsymbol X})$ is similarly obtained as
\begin{align}
    \dot{r} &=-KR_0^2 r+\text{grad}_{\boldsymbol{X}= \boldsymbol{X}(t)}\Gamma(\boldsymbol{X})\cdot 
    \boldsymbol{G}(\boldsymbol{X}(t)) p(t)\notag\\
    &\simeq-KR_0^2 r+\text{grad}_{\boldsymbol{X}= \boldsymbol{\chi}(\varphi(t))}\Gamma(\boldsymbol{X})\cdot 
    \boldsymbol{G}\left(\boldsymbol{\chi}(\varphi(t))\right) p(t)\notag\\
    &=-KR_0^2 r+\boldsymbol{I}(\varphi(t))\cdot \boldsymbol{G}(\varphi(t)) p(t),
    \label{eq:amplitude_reduction}
\end{align} 
where $\boldsymbol{I}(\varphi)=\text{grad}_{\boldsymbol{X}= \boldsymbol{\chi}(\varphi)}\Gamma(\boldsymbol{X})$ is the amplitude sensitivity function (a.k.a. isostable response function)~\cite{Mauroy_2013,Wilson_2016,wilson2018greater}.

For the Stuart-Landau oscillator, the phase and amplitude sensitivity functions can be explicitly calculated as
\begin{align}
    \boldsymbol{Z}(\varphi)&=\frac{1}{R_0}(-\sin \varphi, \cos \varphi),
    \label{eq:phase_sensitivity_function}\\
    \boldsymbol{I}(\varphi)&=KR_0(\cos \varphi,~\sin \varphi).
    \label{eq:amplitude_sensitivity_function}
\end{align}
Then, Eqs.~(\ref{eq:phase_reduction}) and (\ref{eq:amplitude_reduction}) are explicitly given as
\begin{align}
    \dot{\varphi}
    &= \omega_0 +\zeta(\varphi)p(t) ,
    \label{eq:dynamics_of_phase_of_ordr_parameter}\\
    \dot{r}
    &=-KR_0^2 r + \kappa(\varphi)p(t),
    \label{eq:dynamics_of_amplitude}
\end{align}
respectively, where we defined the collective phase and amplitude sensitivity functions as
\begin{align}
    \zeta(\varphi) &= \frac{1}{2}\frac{R_0^2+1}{R_0}\sin\varphi,\\
    \kappa(\varphi) &=  \frac{KR_0}{2}(R_0^2-1)\cos\varphi.
    \label{eq:effective_amplitude_sensitivity_function}
\end{align}
Equation~(\ref{eq:dynamics_of_phase_of_ordr_parameter}) describes the collective phase dynamics of the system of coupled oscillators under the external input, which will be used to design the control input, 
and Eq.~(\ref{eq:dynamics_of_amplitude}) describes the deviation from the unperturbed collective oscillations, which will be used  to evaluate the effect of the external input on the mutual synchrony in the numerical simulation.

\section{Optimal control}

In this section, we design the control input $u(t)$ 
to quickly resynchronize the system with the periodic input after the phase shift disturbance as given in Eq.~(\ref{eq:Jet_lag}).
We use the reduced phase dynamics for the complex mean field, Eq.~(\ref{eq:dynamics_of_phase_of_ordr_parameter}), instead of the original high-dimensional Kuramoto model, Eq.~(\ref{eq:Kuramoto_model_external_control}), to design the control input.

We formulate the optimal control problem~\cite{Lewis_2012, Kirk_2004} for the phase difference 
$\hat{\Psi}(t)= \varphi(t) - \Theta(t)$ as follows:
\begin{align}
       &\min_{\hat{\Psi}(t),~u(t)}~~J = \eta\left(\hat{\Psi}(t_f)\right)+
       \int_{t_s}^{t_f} L\left(\hat{\Psi}(t),u(t)\right) dt, \label{eq:cost_function}\\
       &\text{s.t.}\notag\\
       &\dot{\hat{\Psi}}= \Delta\omega
       +\zeta(\hat{\Psi}(t)+\Theta(t))\left(p_0 \sin \Theta(t)+u(t)\right),\notag\\
       &\hspace{2.7mm}\eqcolon f\left(\hat{\Psi}(t),\Theta(t), u(t)\right),
       \label{eq:model_of_state} \\
       &\hat{\Psi}\left(t_s\right)=\varphi\left(t_s\right)-\Theta\left(t_s\right)=
       \Phi\left(t_s\right)-\Theta\left(t_s\right), \label{eq:initial_condition} \\
       &|p(t)| = |p_0\sin\Theta(t)+u(t)|\leq p_0, \label{eq:inequality_constraint} 
\end{align}
where $\Delta\omega=\omega_0-\omega_p$.
Here, the initial time of the control is taken as $t_s$ at which the phase shift is applied to $\Theta(t)$, and $t_f$ is the terminal time of the control, which is set as $t_f=t_s+2T_p$ where $T_p$ is the period of the 
original periodic 
input $p_0\sin\Theta(t)$.
The constraints are as follows: 
Eq.~(\ref{eq:model_of_state}) gives the dynamics of $\hat{\Psi}$,
Eq.~(\ref{eq:initial_condition}) is the initial condition,
and Eq.~(\ref{eq:inequality_constraint}) is the condition on the range of the control input,
that is, the magnitude of the net external input including $u(t)$ does not
exceed the magnitude of the original periodic input $p_0$.

The cost function $J$ consists of the terminal cost and the stage cost,
both of which are assumed to be 
quadratic functions of the sine of the phase difference from the target, i.e.,
\begin{align}
\eta\left(\hat{\Psi}(t_f)\right)&={S_f}\sin^2\left( {\frac{\hat{\Psi}(t_f) - {\Psi}_0}{2}}\right),\\
L\left(\hat{\Psi}(t),u(t)\right)&= Q\sin^2\left( {\frac{\hat{\Psi}(t) - {\Psi}_0}{2}}
       \right)+  \frac{W}{2}{u(t)}^2 ,
\end{align}
where $\Psi_0$ is the target phase difference, 
$S_f>0$ is the weight of the terminal cost associated with the phase difference, 
$Q>0$ is the weight of the stage cost of the phase difference, 
and $W>0$ is the weight of the stage cost of the control input.

To handle the inequality constraint on the range of the input, we use the penalty function method. 
We introduce the penalty function as
\begin{align}
    P(\Theta(t),u(t))=\frac{1}{2}\sum_{l=1}^{2}\left(\max\left\{g_l(\Theta(t),u(t)),0\right\}\right)^2,
    \label{eq:penalty_function}
\end{align}
where $g_1(\Theta(t),u(t))=p_0\left(\sin\Theta(t)-1\right)+u(t)$ and 
$g_2(\Theta(t),u(t))=-p_0\left(\sin\Theta(t)+1\right)-u(t)$.
Introducing the Lagrange multiplier $\lambda(t)$, 
the total cost function including the constraints and the penalty is given by
\begin{align}
    \bar{J} =J
    + \int_{t_s}^{t_f} \Big\{\lambda(t)\left(f(\hat{\Psi}(t),\Theta(t), u(t))-\dot{\hat{\Psi}}(t)\right)\notag\\
    +BP\left(\Theta(t),u(t)\right)\Big\} d t ,
 \end{align}
where $B>0$ is the weight of the penalty.

The control Hamiltonian of the above optimal control problem is given by
\begin{align}
     H&\left(\hat{\Psi}(t),\Theta(t), u(t),\lambda(t)\right)\notag\\
     &=L\left(\hat{\Psi}(t),u(t)\right)+\lambda(t)f\left(\hat{\Psi}(t),\Theta(t), u(t)\right)\notag\\
     &\quad+BP(\Theta(t),u(t)),
\end{align}
and, from the extremum conditions for $\bar{J}$, we obtain the following set of equations;
\begin{align}
     &\dot{\hat{\Psi}}=f\left(\hat{\Psi}(t),\Theta(t), u(t)\right),\\
     &\hat{\Psi}\left(t_s\right)=\varphi\left(t_s\right)-\Theta\left(t_s\right)
     =\Phi\left(t_s\right)-\Theta\left(t_s\right), \\
     &\dot{\lambda}=-\frac{\partial}{\partial \hat{\Psi}}H\left(\hat{\Psi}(t),\Theta(t), u(t),\lambda(t)\right),\\
     &\lambda\left(t_f\right)=\frac{d}{d \hat{\Psi}}\eta\left(\hat{\Psi}(t_f)\right), \\
     &\frac{\partial}{\partial u}H\left(\hat{\Psi}(t),\Theta(t), u(t),\lambda(t)\right)=0.
     \label{eq:control_input}
\end{align}
Solving this two-point boundary value problem numerically, the control input $u(t)$ is obtained from Eq.~(\ref{eq:control_input}).

\section{Numerical simulations}

In this section, we numerically solve the optimal control problem and obtain the optimal control input.
We solve the two-point boundary value problem for $\hat{\Psi}(t)$ and $\lambda(t)$ in the previous section by the shooting method~\cite{Russell_1995}.
We then apply the obtained control input as the feedforward control signal to the original Kuramoto model with finite $N$ and confirm the effectiveness of the control input designed based on the reduced phase model.

\begin{figure}[tb]
    \centering
    \includegraphics[width=1.0\linewidth]{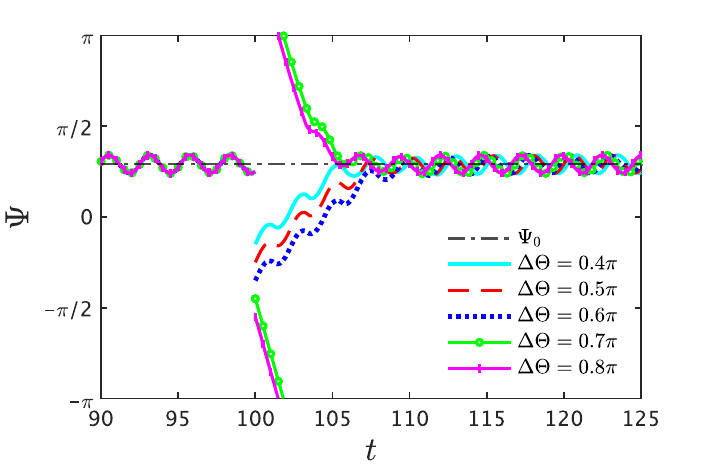}
        \caption{
	Time series of the phase difference $\Psi$ under the optimal control input. The parameters of the Kuramoto model are the same as in Fig.~\ref{fig:Phase_Different_nocont}. The parameters of the optimal control problem are $S_f=1.0$, $Q=5.0$, $W=1.0$, and $B=10^5$.}
    \label{fig:Phase_Different_shooting}
\end{figure}

Figure~\ref{fig:Phase_Different_shooting} shows the time series of the phase difference $\Psi = \Phi -\Theta$ 
under the optimal control.
Comparing with Fig.~\ref{fig:Phase_Different_nocont}, we can clearly observe that the phase locking is reestablished considerably earlier than the case without control.
Table~\ref{tab:recovery_time_opt} shows the recovery time $t_r-t_s$ without and with the control input for several values of the initial phase shift, $\Delta \Theta=0.4\pi,~0.5\pi,~0.6\pi,~0.7\pi$, and $0.8\pi$, applied at $t_s=100$.
Under the optimal control input, resynchronization with the periodic input takes on average 50.28\% of the time of the uncontrolled case.
Thus, the optimal control is useful in reestablishing phase locking from loss of synchronization with the periodic input. 

\begin{figure}[tb]
    \centering
    \includegraphics[width=1.0\linewidth]{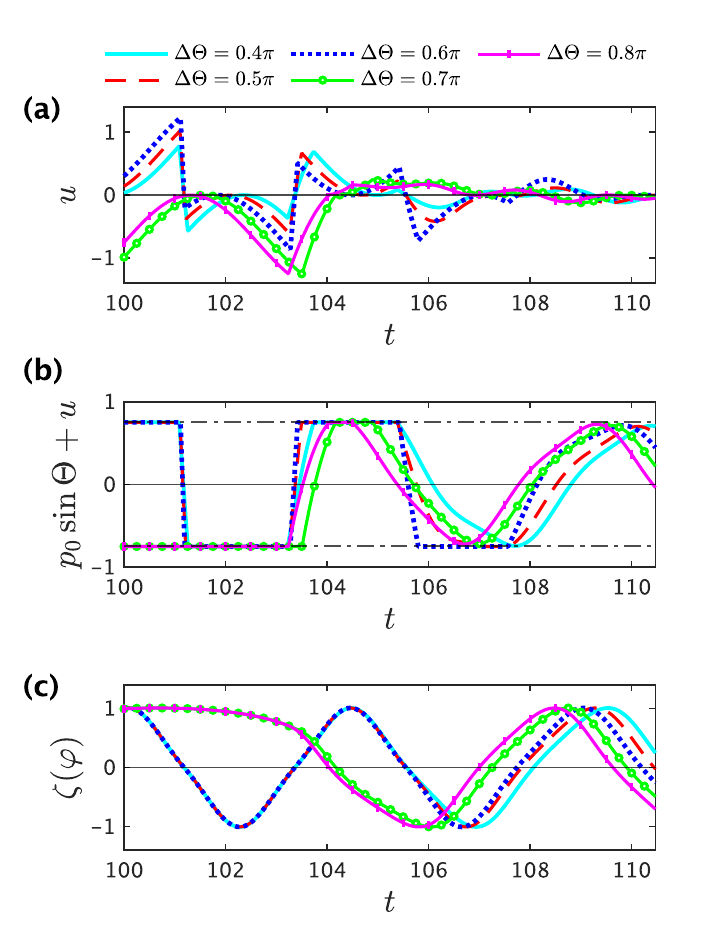}
    \caption{(a) Optimal control input for different values of the phase shift $\Delta \Theta$. 
    (b) Sum of the periodic input and the optimal control input.
    The dash-dotted lines represent $\pm p_0$. (c) Time series of the collective phase sensitivity function.
    }
    \label{fig:input_shooting}
\end{figure}

\begin{table}[tb]
    \centering
    \resizebox{\linewidth}{!}{
    \begin{tabular}{c|c|c|c|c|c}
        \hline
        $\Delta \Theta$ & $0.4\pi$ & $0.5\pi$ & $0.6\pi$ & $0.7\pi$ & $0.8\pi$ \\ \hline
        No   & 12.7    & 17.8    & 16.0    & 13.2    & 10.5    \\
        control  & (2.43$T_p$)   & (3.41$T_p$)   & (3.06$T_p$)   & (2.52$T_p$)   & (2.01$T_p$)   \\\hline
        With  & 7.25     & 7.15     & 9.46    & 5.82     & 5.38   \\ 
        control  & (1.39$T_p$)   & (1.37$T_p$)   & (1.81$T_p$)   & (1.11$T_p$)   & (1.03$T_p$)   \\\hline
    \end{tabular}
     }
    \caption{
    Recovery time $t_r-t_s$ for different values of $\Delta\Theta$.  $T_p=2\pi/\omega_p$ is the period of the input.}
    \label{tab:recovery_time_opt}
\end{table}

Figure~\ref{fig:input_shooting} shows the optimal control input $u(t)$, the sum of the periodic input 
and the control input $p(t)=p_0\sin\Theta(t)+u(t)$, and the collective phase sensitivity function 
$\zeta(\varphi(t))$, respectively. 
We can confirm that the constraint on the range of the control input, Eq.~(\ref{eq:inequality_constraint}), is satisfied.
When $\zeta(\varphi(t))>0$, the phase $\varphi(t)$ is advanced (delayed) for a positive (negative) input,  
and vice versa for $\zeta(\varphi(t))<0$.
For the values of the phase shift $\Delta \Theta=0.4\pi,~0.5\pi,$ and $0.6\pi$, the function $\zeta(\varphi(t))$ and the net external input generally take the same sign, so that the resynchronized is realized by advancing the phase $\varphi(t)$.
On the other hand,  for $\Delta \Theta=0.7\pi$ and $~0.8\pi$, they have generally different signs and the resynchronization occurs by delaying the phase $\varphi(t)$.

\begin{figure}[tb]
    \centering
    \includegraphics[width=1.0\linewidth]{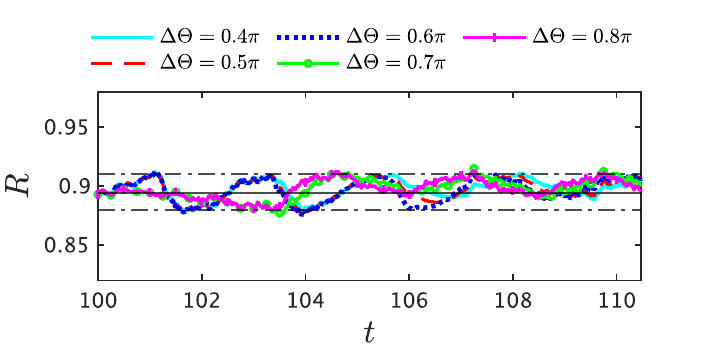}
    \caption{Time series of $R$ with the optimal control input for different values of the phase shift $\Delta \Theta$.
    The black solid line represents $R_0$ and 
    the black dash-dotted lines represent $R_+$ and $R_-$, respectively.}
    \label{fig:R}
\end{figure}

The control input $u(t)$ also affects the degree of mutual synchrony, i.e., the value of $R(t)$, while it may be undesirable to disrupt the mutual synchrony by the control.
We can use the amplitude equation to evaluate the effect of $u(t)$ on $R(t)$.
We here derive the sufficient condition on $r(t)$ for the amplitude decay $-KR_0^2r$ to dominate the dynamics of $r(t)$ within linear approximation.
That is, we seek the condition that
\begin{align}
    KR_0^2|r(t)| \geq |\kappa(\varphi(t)) p(t)|
    \label{eq:condition_r}
\end{align}
holds in Eq.~(\ref{eq:dynamics_of_amplitude})
for any $\kappa(\varphi(t))$ in Eq.~(\ref{eq:effective_amplitude_sensitivity_function}) and 
$p(t)$ in Eq.~(\ref{eq:inequality_constraint}).
If Eq.~(\ref{eq:effective_amplitude_sensitivity_function}), Eq.~(\ref{eq:inequality_constraint}), and $R_0\leq1$ are satisfied, then 
\begin{align}
    \frac{KR_0p_0}{2}(1-R_0^2) \geq |\kappa(\varphi(t))p(t)|
    \label{eq:for_any}
\end{align}
holds.
Thus, the sufficient condition on $r(t)$ to satisfy Eq.~(\ref{eq:condition_r}) 
for any $\kappa(\varphi(t))$ in Eq.~(\ref{eq:effective_amplitude_sensitivity_function}) and 
$p(t)$ in Eq.~(\ref{eq:inequality_constraint}) is given by
\begin{align}
    KR_0^2|r(t)| &\geq \frac{KR_0p_0}{2}(1-R_0^2),
    \label{eq:condition_r_for_any}
\end{align}
i.e.,
\begin{align}
    |r(t)| &\geq  \frac{p_0}{2R_0}(1-R_0^2).
\end{align}
As long as $r(t)$ satisfies Eq.~(\ref{eq:condition_r_for_any}), the system state approaches the limit cycle even
under the external input $p(t)$ containing the control input $u(t)$ 
because the decay term $-KR_0^2r$ dominates.
Here, the condition Eq.~(\ref{eq:condition_r_for_any}) for the original amplitude 
$R = \sqrt{X^2+Y^2}$ is obtained by using the amplitude function in Eq.~(\ref{eq:amplitude_function}) as
\begin{align}
    \left|1-\frac{R_0^2}{\left(R(t)\right)^2}\right|\geq
    \frac{p_0}{KR_0}(1-R_0^2).
\end{align} 
The above condition can be expressed as
\begin{align}
    R(t)&\geq R_0\sqrt{\frac{KR_0}{-(1-R_0^2)p_0+KR_0}}\eqcolon R_+,
    \label{eq:Rp}\\
    R(t)&\leq R_0\sqrt{\frac{KR_0}{(1-R_0^2)p_0+KR_0}}\eqcolon R_-,
    \label{eq:Rm}
\end{align}
where we used $-(1-R_0^2)p_0+KR_0>0$ in our parameter setting. 
Thus, if $R$ satisfies Eqs.~(\ref{eq:Rp}) and (\ref{eq:Rm}), the system state always approaches the limit cycle,
and the system state is expected to stay within a certain distance from the limit cycle, i.e.
between $R_+$ and $R_-$.
Figure~\ref{fig:R} shows the time series of $R$ and the three constants $R_0$, $R_{+}$ and $R_-$.
We can observe that $R(t)$ mostly stays between $R_+$ and $R_-$, and remains near the limit cycle $R_0$.
That is, the control input gives only a limited effect on the mutual synchrony and does not degrade it.
Thus, the effect of the control input  on the degree of mutual synchrony can be estimated by using the amplitude equation.

\section{Conclusion}

We presented a framework to control the synchronization of the high-dimensional Kuramoto model with 
a periodic input by applying the optimal control theory to the low-dimensional equation derived by 
dynamical reduction methods, i.e., the Ott-Antonsen ansatz and phase-amplitude reduction.
Considering a jet lag like situation, we set up a control problem for the system to recover quickly from the loss of synchronization with the periodic input, and demonstrated the effectiveness of the proposed framework by numerical simulations.
In addition, we showed that the control input did not degrade the degree of mutual synchrony by evaluating 
the effect of the input using the amplitude equation.

In this study, we treated the simple Kuramoto model as the controlled object.
However, it may not fully capture synchronization in real-world systems.
Many studies have been conducted to extend the Kuramoto model to reproduce more complex dynamics: 
for example, the cases where the coupling strength depends on time~\cite{Petkoski_2012}, the cases involving higher-order interactions determined by relationships among three or more oscillators~\cite{Skardal_2020, León_2024}, 
and the cases where interactions include time delays~\cite{Yeung_1999}. 
Considering these studies, it is desirable to expand our control framework to more complex systems of coupled oscillators.

Our control framework in the present study is feedforward and calculates the control input beforehand.
Since the dynamical reduction approach can significantly shorten the computation time, 
it is also well suited to real-time control methods, such as model predictive control, 
which solve the optimal control problem at each time step in real time.
By combining the reduced dynamics with model predictive control, it may be possible to control 
high-dimensional systems even in more complex situations, such as trajectory tracking control.

\section{Acknowledgment}
We appreciate N. Namura and K. Taga for productive discussions.
This work was supported by JST SPRING, Japan Grant Number JPMJSP2180,
and by JSPS KAKENHI 25K03081 and 22H00516.


\begin{thebibliography}{10}


\bibitem{Winfree_1980}
Arthur~T. Winfree.
\newblock {\em The Geometry of Biological Time}.
\newblock Springer-Verlag, Berlin, Germany, 1980.

\bibitem{glass1988clocks}
Leon Glass and Michael~C Mackey.
\newblock {\em From clocks to chaos: The rhythms of life}.
\newblock Princeton University Press, 1988.

\bibitem{Strogatz_1993}
Steven~H. Strogatz and Ian Stewart.
\newblock Coupled oscillators and biological synchronization.
\newblock {\em Scientific American}, 269(6):102--109, 1993.

\bibitem{stefanovska1999physics}
Aneta Stefanovska.
\newblock Physics of the human cardiovascular system.
\newblock {\em Contemporary Physics}, 40(1):31--55, 1999.

\bibitem{Pikovsky_2001}
Arkady Pikovsky, Michael Rosenblum, and Jürgen Kurths.
\newblock {\em Synchronization: A Universal Concept in Nonlinear Sciences}.
\newblock Cambridge Nonlinear Science Series. Cambridge University Press, 2001.

\bibitem{Kuramoto_2003}
Yoshiki Kuramoto.
\newblock {\em Chemical oscillations, waves, and turbulence}.
\newblock Chemistry Series. Dover Publications, 2003.
\newblock originally published: Springer Berlin, New York, Heidelberg, 1984.

\bibitem{ermentrout2010mathematical}
Bard Ermentrout and David~Hillel Terman.
\newblock {\em Mathematical foundations of neuroscience}, volume~35.
\newblock Springer, 2010.

\bibitem{Yamaguchi_2003}
Shun Yamaguchi, Hiromi Isejima, Takuya Matsuo, Ryusuke Okura, Kazuhiro Yagita, Masaki Kobayashi, and Hitoshi Okamura.
\newblock Synchronization of cellular clocks in the suprachiasmatic nucleus.
\newblock {\em Science}, 302(5649):1408--1412, 2003.

\bibitem{Golombek_2010}
Diego~A. Golombek and Ruth~E. Rosenstein.
\newblock Physiology of circadian entrainment.
\newblock {\em Physiological Reviews}, 90(3):1063--1102, 2010.
\newblock PMID: 20664079.

\bibitem{Karma_2013}
Alain Karma.
\newblock Physics of cardiac arrhythmogenesis.
\newblock {\em Annual Review of Condensed Matter Physics}, 4(Volume 4, 2013):313--337, 2013.

\bibitem{Dirk_2012}
Martin Rohden, Andreas Sorge, Marc Timme, and Dirk Witthaut.
\newblock Self-organized synchronization in decentralized power grids.
\newblock {\em Phys. Rev. Lett.}, 109:064101, Aug 2012.

\bibitem{Marco_2009}
Anders~Lyhne Christensen, Rehan OGrady, and Marco Dorigo.
\newblock From fireflies to fault-tolerant swarms of robots.
\newblock {\em IEEE Transactions on Evolutionary Computation}, 13(4):754--766, 2009.

\bibitem{Moehlis_2006a}
Jeff Moehlis.
\newblock Canards for a reduction of the {H}odgkin-{H}uxley equations.
\newblock {\em Journal of Mathematical Biology}, 52:141--153, 2006.

\bibitem{Moehlis_2006b}
Jeff Moehlis, Eric Shea-Brown, and Herschel Rabitz.
\newblock Optimal inputs for phase models of spiking neurons.
\newblock {\em Journal of Computational and Nonlinear Dynamics}, 1(4):358--367, 06 2006.

\bibitem{Zlotnik_2013}
Anatoly Zlotnik, Yifei Chen, Istv{\'a}n~Z Kiss, Hisa-Aki Tanaka, and Jr-Shin Li.
\newblock Optimal waveform for fast entrainment of weakly forced nonlinear oscillators.
\newblock {\em Phys. Rev. Lett.}, 111(2):024102, 2013.

\bibitem{Jr-Shin_2016}
Anatoly Zlotnik, Raphael Nagao, István Kiss, and Jr-Shin Li.
\newblock Phase-selective entrainment of nonlinear oscillator ensembles.
\newblock {\em Nature Communications}, 7:10788, 03 2016.

\bibitem{Tanaka_2008}
Takuma Tanaka and Toshio Aoyagi.
\newblock Optimal weighted networks of phase oscillators for synchronization.
\newblock {\em Phys. Rev. E}, 78:046210, Oct 2008.

\bibitem{Kawamura_2008}
Yoji Kawamura, Hiroya Nakao, Kensuke Arai, Hiroshi Kori, and Yoshiki Kuramoto.
\newblock Collective phase sensitivity.
\newblock {\em Phys. Rev. Lett.}, 101:024101, Jul 2008.

\bibitem{Monga_2019}
Bharat Monga, Dan Wilson, Tim Matchen, and Jeff Moehlis.
\newblock Phase reduction and phase-based optimal control for biological systems: a tutorial.
\newblock {\em Biol. Cybern}, 113(1-2):11--46, Apr 2019.

\bibitem{Takata_2021}
Shohei Takata, Yuzuru Kato, and Hiroya Nakao.
\newblock Fast optimal entrainment of limit-cycle oscillators by strong periodic inputs via phase-amplitude reduction and floquet theory.
\newblock {\em Chaos: An Interdisciplinary Journal of Nonlinear Science}, 31(9):093124, 09 2021.

\bibitem{Kato_2021}
Yuzuru Kato, Anatoly Zlotnik, Jr-Shin Li, and Hiroya Nakao.
\newblock Optimization of periodic input waveforms for global entrainment of weakly forced limit-cycle oscillators.
\newblock {\em Nonlinear Dynamics}, 105(3):2247--2263, 2021.

\bibitem{Ozawa_2021}
Ayumi Ozawa and Hiroshi Kori.
\newblock Feedback-induced desynchronization and oscillation quenching in a population of globally coupled oscillators.
\newblock {\em Phys. Rev. E}, 103:062217, Jun 2021.

\bibitem{Berman_2022}
Carlo Sinigaglia, Francesco Braghin, and Spring Berman.
\newblock Optimal control of velocity and nonlocal interactions in the mean-field kuramoto model.
\newblock In {\em 2022 American Control Conference (ACC)}, pages 290--295, 2022.

\bibitem{Petar_2023}
Petar Mircheski, Jinjie Zhu, and Hiroya Nakao.
\newblock Phase-amplitude reduction and optimal phase locking of collectively oscillating networks.
\newblock {\em Chaos: An Interdisciplinary Journal of Nonlinear Science}, 33(10):103111, 10 2023.

\bibitem{Yawata_2024}
Koichiro Yawata, Kai Fukami, Kunihiko Taira, and Hiroya Nakao.
\newblock Phase autoencoder for limit-cycle oscillators.
\newblock {\em Chaos: An Interdisciplinary Journal of Nonlinear Science}, 34(6):063111, 06 2024.

\bibitem{Wilson_2024}
Adharaa~Neelim Dewanjee and Dan Wilson.
\newblock Optimal phase-based control of strongly perturbed limit cycle oscillators using phase reduction techniques.
\newblock {\em Phys. Rev. E}, 109:024223, Feb 2024.

\bibitem{Namura_2024a}
Norihisa Namura and Hiroya Nakao.
\newblock Optimal coupling functions for fast and global synchronization of weakly coupled limit-cycle oscillators.
\newblock {\em Chaos, Solitons \& Fractals}, 185:115080, 2024.

\bibitem{Namura_2024b}
Norihisa Namura and Hiroya Nakao.
\newblock Optimal phase control of limit-cycle oscillators with strong inputs through phase-amplitude reduction.
\newblock In {\em 2024 IEEE 63rd Conference on Decision and Control (CDC)}, pages 4003--4009, 2024.

\bibitem{Strogatz_2000}
Steven~H. Strogatz.
\newblock From {K}uramoto to {C}rawford: exploring the onset of synchronization in populations of coupled oscillators.
\newblock {\em Physica D: Nonlinear Phenomena}, 143(1):1--20, 2000.

\bibitem{Ott_Antonsen_2008}
Edward Ott and Thomas~M. Antonsen.
\newblock Low dimensional behavior of large systems of globally coupled oscillators.
\newblock {\em Chaos: An Interdisciplinary Journal of Nonlinear Science}, 18(3), September 2008.

\bibitem{Kawamura_2010a}
Yoji Kawamura, Hiroya Nakao, Kensuke Arai, Hiroshi Kori, and Yoshiki Kuramoto.
\newblock Phase synchronization between collective rhythms of globally coupled oscillator groups: Noisy identical case.
\newblock {\em Chaos: An Interdisciplinary Journal of Nonlinear Science}, 20(4):043109, 11 2010.

\bibitem{Kawamura_2010b}
Yoji Kawamura, Hiroya Nakao, Kensuke Arai, Hiroshi Kori, and Yoshiki Kuramoto.
\newblock Phase synchronization between collective rhythms of globally coupled oscillator groups: Noiseless nonidentical case.
\newblock {\em Chaos: An Interdisciplinary Journal of Nonlinear Science}, 20(4):043110, 11 2010.

\bibitem{Wolfrum_2013}
Oleh~E. Omel’chenko and Matthias Wolfrum.
\newblock Bifurcations in the sakaguchi–kuramoto model.
\newblock {\em Physica D: Nonlinear Phenomena}, 263:74--85, 2013.

\bibitem{Strogatz_2008}
Lauren~M. Childs and Steven~H. Strogatz.
\newblock Stability diagram for the forced kuramoto model.
\newblock {\em Chaos: An Interdisciplinary Journal of Nonlinear Science}, 18(4):043128, 12 2008.

\bibitem{Ernest_2014}
Diego Paz\'o and Ernest Montbri\'o.
\newblock Low-dimensional dynamics of populations of pulse-coupled oscillators.
\newblock {\em Phys. Rev. X}, 4:011009, Jan 2014.

\bibitem{Pikovsky_2019}
Chen~Chris Gong, Chunming Zheng, Ralf Toenjes, and Arkady Pikovsky.
\newblock Repulsively coupled kuramoto-sakaguchi phase oscillators ensemble subject to common noise.
\newblock {\em Chaos: An Interdisciplinary Journal of Nonlinear Science}, 29(3):033127, 03 2019.

\bibitem{Lewis_2012}
Frank~L Lewis, Draguna Vrabie, and Vassilis~L Syrmos.
\newblock {\em Optimal control}.
\newblock John Wiley \& Sons, 2012.

\bibitem{Kirk_2004}
Donald~E. Kirk.
\newblock {\em Optimal Control Theory: An Introduction}.
\newblock Dover Books on Electrical Engineering Series. Dover Publications, 2004.

\bibitem{Wilson_2016}
Dan Wilson and Jeff Moehlis.
\newblock Isostable reduction of periodic orbits.
\newblock {\em Phys. Rev. E}, 94:052213, Nov 2016.

\bibitem{Shirasaka_2017}
Sho Shirasaka, Wataru Kurebayashi, and Hiroya Nakao.
\newblock Phase-amplitude reduction of transient dynamics far from attractors for limit-cycling systems.
\newblock {\em Chaos: An Interdisciplinary Journal of Nonlinear Science}, 27(2):023119, 02 2017.

\bibitem{Nakao_2016}
Hiroya Nakao.
\newblock Phase reduction approach to synchronisation of nonlinear oscillators.
\newblock {\em Contemporary Physics}, 57(2):188--214, 2016.

\bibitem{Kori_2017}
Hiroshi Kori, Yoshiaki Yamaguchi, and Hitoshi Okamura.
\newblock Accelerating recovery from jet lag: prediction from a multi-oscillator model and its experimental confirmation in model animals.
\newblock {\em Scientific Reports}, 7(1):46702, 2017.

\bibitem{Lu_2016}
Zhixin Lu, Kevin Klein-Carde{\~n}a, Steven Lee, Thomas~M Antonsen, Michelle Girvan, and Edward Ott.
\newblock Resynchronization of circadian oscillators and the east-west asymmetry of jet-lag.
\newblock {\em Chaos}, 26(9):094811, Sep 2016.

\bibitem{Acebron_2005}
Juan~A. Acebr\'on, L.~L. Bonilla, Conrad~J. P\'erez~Vicente, F\'elix Ritort, and Renato Spigler.
\newblock The kuramoto model: A simple paradigm for synchronization phenomena.
\newblock {\em Rev. Mod. Phys.}, 77:137--185, Apr 2005.

\bibitem{Mauroy_2013}
A.~Mauroy, I.~Mezić, and J.~Moehlis.
\newblock Isostables, isochrons, and {K}oopman spectrum for the action–angle representation of stable fixed point dynamics.
\newblock {\em Physica D: Nonlinear Phenomena}, 261:19--30, 2013.

\bibitem{wilson2018greater}
Dan Wilson and Bard Ermentrout.
\newblock Greater accuracy and broadened applicability of phase reduction using isostable coordinates.
\newblock {\em Journal of Mathematical Biology}, 76:37--66, 2018.

\bibitem{Russell_1995}
Uri~M. Ascher, Robert M.~M. Mattheij, and Robert~D. Russell.
\newblock {\em Numerical Solution of Boundary Value Problems for Ordinary Differential Equations}.
\newblock Society for Industrial and Applied Mathematics, 1995.

\bibitem{Petkoski_2012}
Spase Petkoski and Aneta Stefanovska.
\newblock Kuramoto model with time-varying parameters.
\newblock {\em Phys. Rev. E}, 86:046212, Oct 2012.

\bibitem{Skardal_2020}
Per~Sebastian Skardal and Alex Arenas.
\newblock Higher order interactions in complex networks of phase oscillators promote abrupt synchronization switching.
\newblock {\em Communications Physics}, 3(1):218, 2020.

\bibitem{León_2024}
Iván León, Riccardo Muolo, Shigefumi Hata, and Hiroya Nakao.
\newblock Higher-order interactions induce anomalous transitions to synchrony.
\newblock {\em Chaos: An Interdisciplinary Journal of Nonlinear Science}, 34(1):013105, 01 2024.

\bibitem{Yeung_1999}
M.~K.~Stephen Yeung and Steven~H. Strogatz.
\newblock Time delay in the {K}uramoto model of coupled oscillators.
\newblock {\em Phys. Rev. Lett.}, 82:648--651, Jan 1999.

\end{thebibliography}

\end{document}